\documentclass[preprint,showpacs,preprintnumbers,amsmath,amssymb]{revtex4}

\usepackage{graphicx}
\usepackage{dcolumn}
\usepackage{bm}
\usepackage{amssymb}

\newcommand{\vecQ}{\mbox{\boldmath $Q$}}

\newcommand{\vecq}{\mbox{\boldmath $q$}}

\begin{document}
\title{The Jarzynski equality in van der Pol and Rayleigh oscillators 
}

\author{Hideo Hasegawa}
\altaffiliation{hideohasegawa@goo.jp;
http://sites.google.com/site/hideohasegawa/}
\affiliation{Department of Physics, Tokyo Gakugei University,  
Koganei, Tokyo 184-8501, Japan}%

\date{\today}

\begin{abstract}
We have studied the Jarzynski equality (JE) in van der Pol and 
Rayleigh oscillators which are typical deterministic non-Hamiltonian models, 
but not expected to rigorously satisfy the JE 
because they are not microscopically reversible. 
Our simulations that calculate the contribution to the work 
$W$ of an applied ramp force with a duration $\tau$ show that
the JE approximately holds for a fairly wide range
of $\tau$ including $\tau \rightarrow 0 $ and $\tau \rightarrow \infty$,
except for $\tau \sim T$ where $T$ denotes the period of relaxation oscillations 
in the limit cycle. 
The work distribution function (WDF) is shown to be non-Gaussian 
with the $U$-shaped structure for a strong damping parameter.
The $\tau$ dependence of $R$ $\left(=- k_B T \ln \left< e^{-\beta W} \right> \right)$ 
obtained by our simulations is semi-quantitatively elucidated
with the use of a simple expression for limit-cycle oscillations,
where the bracket $\langle \cdot \rangle$ expresses an average over the WDF.
The result obtained in self-excited oscillators is in contrast with the fact
that the JE holds in the Nos\'{e}-Hoover oscillator 
which also belongs to deterministic non-Hamiltonian models.

\end{abstract}

        


\maketitle
\newpage
\section{Introduction}

In the last decade, a significant progress has been made
in theoretical studies on nonequilibrium statistics
(for reviews, see Refs. \cite{Busta05,Ritort07,Ciliberto10}). 
The important three fluctuation theorems have been proposed:
the Jarzynski equality (JE) \cite{Jarzynski97}, 
the steady-state and transient fluctuation theorems 
\cite{Evans93,Evans94,Crooks99,Narayan04},
and the Crooks theorem \cite{Crooks99,Narayan04}.
They may be applicable to nonequilibrium systems
driven arbitrarily far from the equilibrium states.
In this paper we pay our attention to the JE expressed by
\begin{eqnarray}
\left< e^{-\beta W} \right> 
&=& \int dW \:P(W) \:e^{-\beta W}=e^{- \beta \Delta F},
\label{eq:A1}
\end{eqnarray}
where $W$ denotes a work made in a system when its parameter 
is changed, the bracket $\langle \cdot \rangle$ expresses
the average over the work distribution function (WDF), $P(W)$,
of a work performed by a prescribed protocol,
$\Delta F$ stands for the free energy difference between the initial and final equilibrium states,
and $\beta$ $(=1/k_B T)$ is the inverse temperature of the initial state.
Equation (\ref{eq:A1}) includes the second law of thermodynamics, 
$\langle W \rangle \geq \Delta F$, 
where the equality holds only for the reversible process.
The JE was originally proposed for classical isolated system and open system weakly coupled to baths 
which are described by the Hamiltonian \cite{Jarzynski97} and the stochastic models 
\cite{Jarzynski97b}.
Jarzynski later proved that the JE is valid for strongly coupled open systems \cite{Jarzynski04}.  
A validity of the JE has been confirmed by some experiments 
for systems which may be described by damped harmonic oscillator models
\cite{Liphardt02,Wang05,Douarche05,Douarche06,Joubaud07,Joubaud07b}.
Stimulated by these experiments, many theoretical analyses have been made for 
harmonic oscillators with the use of the Markovian Langevin model 
\cite{Douarche05,Douarche06,Joubaud07,Joubaud07b},
the non-Markovian Langevin model \cite{Zamponi05,Mai07,Speck07,Ohkuma07},
Fokker-Planck equation \cite{Chaudhury08}, and Hamiltonian model 
\cite{Jarzynski06,Jarzynski08,Dhar05,Chakrabarti08,Hijar10,Hasegawa11b}.
Recently the validity of the JE in nonlinear oscillators with anharmonic potentials
has been investigated in Refs. \cite{Mai07,Saha06}.

In this paper, we will study the self-excited oscillators described by
van der Pol and Rayleigh equations with state- and velocity-dependent dampings
\cite{Pol26,Rayleigh83}. They are expressed by
\begin{eqnarray}
\dot{x} &=& v, 
\label{eq:G1} \\
\dot{v} &=& - x - \zeta \:v + f(t), 
\label{eq:G2} \\
\zeta &=& \left\{ \begin{array}{ll}
c(x^2-a)
\quad & \mbox{for the van der Pol model}, 
\label{eq:G4}\\ 
c(v^2-a)
\quad & \mbox{for the Rayleigh model},
\label{eq:G5}
\end{array} \right. 
\end{eqnarray}
where $v=\dot{x}$, $a=1$, $c$ ($\geq 0$) is a damping parameter,  
dot $(\cdot)$ denotes a derivative with respect to time $t$,
$\zeta$ stands for an auxiliary variable, and
$f(t)$ expresses an applied external force.
Conditions of $\zeta > 0$ and $\zeta < 0$ express 
positive and negative dissipations, respectively.
The van der Pol equation was proposed as a mathematical model of self-excited oscillations 
for a simple electric circuit with nonlinear triode valve \cite{Pol26}. 
The Rayleigh equation was introduced to show the appearance of sustained
vibrations in acoustics \cite{Rayleigh83}.
Van der Pol and Rayleigh equations 
are formally equivalent in the sense that the van der Pol equation may be 
transformed to the Rayleigh equation  and vice versa with a proper change of variables.
These equations provide basic models for various nonlinear dynamics
of systems in mechanical and electrical engineerings, biology, biochemistry
and many other applications 
(for a recent review on nonlinear equations, see Ref. \cite{He06}).
Many studies have been reported on the van der Pol model,
which may be regarded as a special case of the FitzHugh-Nagumo model
\cite{Fitz61,Nagumo62}. 
The properties of periodic solutions of the van der Pol oscillator
are known in considerable detail
for a sufficiently small or large damping coefficient.

The van der Pol and Rayleigh oscillators belong to deterministic non-Hamiltonian
models. The Nos\'{e}-Hoover (NH) oscillator \cite{Nose84,Hoover85} which
has been widely adopted for a study of molecular dynamics also belongs 
to non-Hamiltonian models.
The NH oscillator is described by \cite{Nose84,Hoover85}
\begin{eqnarray}
\dot{x} &=& v, 
\label{eq:G1b} \\
\dot{v} &=& - x - \zeta \:v + f(t), 
\label{eq:G2b} \\
\dot{\zeta} &=& \frac{1}{\tau_Q}(v^2-k_B T),
\label{eq:G3}
\end{eqnarray}
where $\zeta$ is a state variable of the thermal reservoir 
with the temperature $T$ and $\tau_Q$ stands for the relaxation time of $\zeta$.
It is noted that Eqs. (\ref{eq:G1})-(\ref{eq:G5}) with $a=k_B T$ are similar 
to Eqs. (\ref{eq:G1b})-(\ref{eq:G3}) except for the fact that
Eq. (\ref{eq:G4}) is given by $\zeta$ while 
Eq. (\ref{eq:G3}) is expressed by $\dot{\zeta}$ \cite{Note2}.
The fluctuation theorem in non-Hamiltonian system coupled to thermostat has been 
discussed in Refs. \cite{Evans93,Evans94,Gallavotti95,Evans95,Sevick08}.
Refs. \cite{Cuendet06,Cuendet06b,Paschinger06} have provided 
the condition for the JE to hold in non-Hamiltonian (and Hamiltonian) model.
The condition requires that 
the equilibrium canonical distribution should be given by \cite{Paschinger06}
\begin{eqnarray}
P(\vecQ, \vecq, \lambda) &\propto& e^{-\beta H\left(\vecQ, \:\lambda \right)} 
\:e^{-\beta \psi(\vecq)},
\label{eq:G6}
\end{eqnarray}
with
\begin{eqnarray}
\psi(\vecq) &=& -k_B T \ln \phi(\vecq),
\label{eq:G6b}
\end{eqnarray}
where $H(\vecQ)$ denotes energy of the system, $\phi(\vecq)$ is the (normalized) 
equilibrium distribution of bath variable $\vecq$ and $\lambda$ is an external parameter.
The canonical distribution of the Nos\'{e}-Hoover model with $f(t)=\lambda$ is
given by \cite{Hoover85}
\begin{eqnarray}
P(x, v, \zeta) &\propto& e^{-\beta(x^2/2- \lambda x +v^2/2+\tau_Q \zeta^2/2)},
\label{eq:G7}
\end{eqnarray}
which satisfies the condition given by Eqs. (\ref{eq:G6}) and (\ref{eq:G6b})
with $\vecQ=(x,v)$ and $\vecq=\zeta$, and then the JE holds 
in the Nos\'{e}-Hoover model \cite{Cuendet06,Cuendet06b,Paschinger06}.
In contrast, the equilibrium distribution of the van der Pol or Rayleigh oscillator
which is an odd-shaped racetrack \cite{Holian89}, does not meet the condition 
given by Eqs. (\ref{eq:G6}) and (\ref{eq:G6b}). 
Although this suggests that the JE does not hold in van der Pol and Rayleigh 
oscillators, it is worthwhile to examine how and to what extent the JE is violated 
in self-excited oscillators, which is the purpose of the present paper.

The paper is organized as follows. In the next Sec. II, we briefly explain basic 
equations of van der Pol and Rayleigh oscillators. 
We examine the validity of the JE by simulations applying
a ramp force with a duration $\tau$.
Our simulations in Sec. III show that 
although the JE is not exactly satisfied in self-excited oscillators, 
the JE approximately holds in a fairly wide range of $\tau$ values
including $\tau \rightarrow 0$ (transient force) and $\tau \rightarrow \infty$  
(quasi-stationary force).
Various types of analytical solutions for applied sinusoidal forces 
have been developed for van der Pol and Rayleigh models.
It is, however, still difficult to obtain analytical solutions 
for arbitrary external forces including non-periodic ones.
By using a simple analytic expression of solutions for an applied ramp force
which is suggested by He's method for a limit cycle of self-excited oscillators \cite{He03},
we present in Sec. IV, a semi-quantitative analysis of the results of our simulations.
Sec. V is devoted to our conclusion.

\section{Self-excited oscillator models}

\subsection{Energy, heat and work}

From Eqs. (\ref{eq:G1})-(\ref{eq:G5}), van der Pol and Rayleigh oscillators 
are described by
\begin{eqnarray}
\ddot{x} + x + \zeta \:v &=& f(t),
\label{eq:A2}
\end{eqnarray}
with
\begin{eqnarray}
\zeta &=& \left\{ \begin{array}{ll}
c(x^2-1)
\quad & \mbox{for the van der Pol model}, \\ 
c(v^2-1)
\quad & \mbox{for the Rayleigh model}.
\end{array} \right. 
\end{eqnarray} 
When we set $\zeta=c$ $(> 0)$ in Eq. (\ref{eq:A2}), it expresses a damped harmonic oscillator.
Multiplying $\dot{x}$ for the both sides of Eq. (\ref{eq:A2}) and
integrating them over $t$, we obtain
\begin{eqnarray}
U(t)-U(0) &=& Q(t)+ W_c(t),
\label{eq:D1}
\end{eqnarray}
with
\begin{eqnarray}
U(t) &=& \frac{\dot{x}(t)^2}{2}+\frac{x(t)^2}{2}, 
\label{eq:D2}\\
Q(t) &=& - \int_{0}^{t} \zeta \:\dot{x}^2 \:dt,
\label{eq:D3}\\
W_c(t) &=& \int_{0}^{t} f(t) \:\dot{x} \:dt,
\label{eq:D4} 
\end{eqnarray}
where $U(t)$, $Q(t)$ and $W_c(t)$ stand for 
the internal energy, heat (dissipative energy) and classical work, respectively.
Equation (\ref{eq:D1}) expresses the first law of thermodynamics.
In order to show the JE, Jarzynski employed an alternative work 
defined by \cite{Jarzynski97}
\begin{eqnarray}
W_J(t) &=& - \int_{0}^{t} \dot{f}(t) \:x(t) \:dt,
\label{eq:A6}
\end{eqnarray}
which is related with $W_c(t)$ as
\begin{eqnarray}
W_J(t) &=& - f(t)x(t)+f(0)x(0) + W_c(t).
\end{eqnarray}
It is noted that $U(t)$, $Q(t)$, $W_c(t)$ and $W_J(t)$ depend on a microscopic 
history of the system of $x(t)$ and $v(t)$ for $t \geq 0$ starting from
their initial values of $x(0)$ $(=x_0)$ and $v(0)$ $(=v_0)$.
$W_J(t)$ has been employed for a study of the JE in this study.

We have presented in the Appendix, some numerical calculations 
of thermodynamical quantities such as energy, heat and work
of the van der Pol oscillator, 
which are evaluated both by single and multiple runs of simulations. 
It should be note that even for $f(t)=0$,
we obtain $\langle U(t) \rangle_0-\langle U(0) \rangle_0 \neq 0$
in van der Pol (and Rayleigh) oscillators 
because of a dissipative contribution of $\langle dQ(t)/dt \rangle_0$
(see Figs. \ref{fig12} and \ref{fig13} in the Appendix), 
where $\langle \cdot \rangle_0$ stands for an average over initial states
[Eqs. (\ref{eq:A8b}) and (\ref{eq:A9})].
This is in contrast to the NH oscillator
where the relations, $\langle U(t) \rangle_0-\langle U(0) \rangle_0 = 0$ 
and  $\langle dQ(t)/dt \rangle_0=0$, hold. 
This difference reflects on the difference in non-equilibrium properties
of self-excited and NH oscillators: the JE does not hold in the former while
it holds in the latter.

\subsection{The Jarzynski equality}
For a study of the JE, we will apply a ramp force $f(t)$ given by
\begin{eqnarray}
f(t) &=& \left\{ \begin{array}{ll}
0
\quad & \mbox{for $t < 0 $}, \\ 
g \left( \frac{t}{\tau} \right)
\quad & \mbox{for $0 \leq t < \tau $}, \\ 
g
\quad & \mbox{for $t \geq \tau $},
\end{array} \right. 
\label{eq:A5}
\end{eqnarray}
where $\tau$ denotes a duration of the force and $g$ its magnitude.
By using Eqs. (\ref{eq:A6}) and (\ref{eq:A5}), we obtain a work induced 
by the applied ramp force,
\begin{eqnarray}
W_0 &\equiv& W_J(\tau)= -\left( \frac{g}{\tau} \right) \int_0^{\tau} x(t) \:dt.
\label{eq:A7}
\end{eqnarray}
The WDF is expressed by
\begin{eqnarray}
P(W) &=& \left<  \delta(W-W_0)\right>_0,
\label{eq:A8}
\end{eqnarray}
where $\langle \cdot \rangle_0$ signifies the average over
the canonically distributed $\{ x_0 \}$ and $\{ v_0 \}$,
\begin{eqnarray}
P(x_0, v_0) &\propto& e^{-\beta (x_0^2/2+v_0^2/2)},
\label{eq:A8b}
\end{eqnarray}
satisfying the equi-partition relation given by 
\begin{eqnarray}
\left< x_0^2 \right>_{0} &=& \left< v_0^2 \right>_{0} = k_B T.
\label{eq:A9}
\end{eqnarray}

The JE in Eq. (\ref{eq:A1}) may be rewritten as
\begin{eqnarray}
R &\equiv& -\frac{1}{\beta} \ln \left< e^{-\beta W} \right>= \Delta F.
\label{eq:A10}
\end{eqnarray}
If the WDF is Gaussian, $R$ is given by
\begin{eqnarray}
R &=& \mu -\frac{\beta  \sigma^2}{2},
\label{eq:A11}
\end{eqnarray}
with
\begin{eqnarray}
\mu &=& \langle W \rangle, 
\label{eq:A12} \\ 
\sigma^2 &=& \langle (W-\mu)^2 \rangle,
\label{eq:A13}
\end{eqnarray}
which stand for mean and variance, respectively, of the WDF.
Of course Eq. (\ref{eq:A11}) is not valid for non-Gaussian WDF.

\section{Model Calculations}
\subsection{The van der Pol oscillator}

Model calculations of the JE for van der Pol and Rayleigh oscillators
will be reported in Sec. IIIA and IIIB, respectively.
We have made simulations, solving Eq. (\ref{eq:A2}) 
by using the Runge-Kutta method with a time step of 0.0001 
for initial states of $\{ x_0 \}$ and $\{ v_0 \}$ 
given by Eqs. (\ref{eq:A8b}) and (\ref{eq:A9}) with $k_B T=1.0$.

In order to get some insight into the van der Pol oscillator, we first show
results without forces [$f(t)=0.0$].
Time courses of $x(t)$ and $v(t)$ for a damping parameter of $c=10.0$
calculated by single runs with $x_0=1.0$ and $v_0=0.0$
are plotted in Figs. \ref{fig1}(a) and \ref{fig1}(b), respectively. 
Time courses of $x(t)$ exhibit the relaxation oscillation with
characteristic sharp periodic jumps. 
A parametric plot of $x(t)$ vs. $v(t)$ in Fig. \ref{fig1}(c) 
shows the limit cycle.
The period of the relaxation oscillation depends on the
magnitude of a damping parameter $c$. The dashed curve of Fig. \ref{fig1}(d) 
expresses the $c$ dependence of period $T$ 
for $f(t)=0.0$, which is increased with increasing $c$. 
When a constant force $f=0.5$ is applied to
the oscillator, its oscillation period is further increased as shown 
by the solid curve in Fig. \ref{fig1}(d). 

Figure \ref{fig2}(a) shows time courses of $x(t)$ when a ramp force with $\tau=100.0$ 
is applied to the van der Pol oscillator for $c=1.0$ with
initial conditions of $x_0=1.0$ and $v_0=0.0$.
The period of the oscillation with the applied ramp force with $g=0.5$ (solid curve) 
is gradually increased compared to that with $g=0.0$ (dashed curve). 
Figure \ref{fig2}(b) shows a similar plot of $x(t)$ for $c=10.0$. 
The period of the oscillation with $c=10.0$ 
is longer than that with $c=1.0$ by a factor of about three.
An applied ramp force with $g=0.5$ and $\tau=100$ induces a work $W_J(t)$ 
whose time course is shown by the chain (solid) curve for $c=1.0$ ($c=10.0$) 
in Fig. \ref{fig2}(c). We obtain $W_0=-0.110$ ($W_0= -0.126$) 
for $c=1.0$ ($c=10.0$) by a single run with initial values of $x_0=1.0$ and $v_0=0.0$.

Calculating $W_0$ in Eq. (\ref{eq:A7}) for given initial states,
we have obtained the WDF, $P(W)$, with the use of Eq. (\ref{eq:A8}), 
whose results for $\tau=0.1$, 1.0, 10.0 and 100.0 
with $c=10.0$ are plotted in Fig. \ref{fig3}.
Although the WDF for $\tau=0.1$ is Gaussian,
those for $\tau=1.0$, 10.0 and 100.0 are non-Gaussian with the U-shaped structure, 
which are quite different from the Gaussian distributions obtained in harmonic oscillators.
Figure \ref{fig4} shows $P(W)$ for various values of $c$ with
a fixed $\tau=1.0$. 
With decreasing the damping parameter from $c=10.0$, the WDF is changed from
the double-peaked distribution to 
the single-peaked Gaussian-like distribution.

Figures  \ref{fig5}(a), \ref{fig5}(b) and \ref{fig5}(c) show $\tau$ dependences of
$\mu$, $\sigma$ and $R$, respectively, 
for $c=1.0$ (dashed curves), $c=5.0$ (dotted curves) and $c=10.0$ (solid curves)
for ramp forces with $g=0.5$.
$\mu$ is almost zero for $\tau=0.1$ and it gradually decreased
to $-0.125$ for $\tau=1000.0$. In contrast, $\sigma \simeq 0.5$ 
at $\tau=0.1$ and it goes to zero at $\tau=1000.0$ with a small bump at $\tau \sim 3.0$.
The calculated $R$ with $c=10.0$ (solid curve)
is in nearly agreement with $\Delta F$ ($=-g^2/2=-0.125$) 
for $\tau \gtrsim 100.0$ and $\tau \lesssim 0.2$,
but it significantly deviates from $\Delta F$ for $0.2 \lesssim \tau \lesssim 100.0$. 
The discrepancy of $R \neq \Delta F$ implies a violation of the JE.
This deviation becomes less significant for smaller values of $c=5.0$ and 1.0, 
and it vanishes for $c=0.0$ (harmonic oscillator) where the JE holds.

\subsection{The Rayleigh oscillator}

Next we study the case of the Rayleigh oscillator.
Time courses of $x(t)$ and $v(t)$
of the relaxation oscillation for $c=10.0$ and $f(t)=0.0$
are plotted in Figs. \ref{fig6}(a) and \ref{fig6}(b), respectively, 
and a parametric plot of $x(t)$ vs. $v(t)$ in Fig. \ref{fig6}(c) exhibits the limit cycle.
The period of the relaxation oscillation depends on the
magnitude of $c$. The dashed curve of Fig. \ref{fig6}(d) 
expresses the $c$ dependence of the period $T$ for $f(t)=0.0$, 
which is increased with increasing $c$. 
The period of the oscillator for a constant $f=0.5$ (solid curve) 
coincides with that for $f=0.0$ (dashed curve) in Fig. \ref{fig6}(d).

The dashed (solid) curve in Fig. \ref{fig7}(a) shows $x(t)$
of the Rayleigh oscillator with $c=1.0$
for ramp forces with $g=0.0$ ($g=0.5$) and $\tau=100.0$, 
which is calculated by single runs with initial condition of $x_0=1.0$ and $v_0=0.0$.
We note that $x(t)$ for $g=0.5$ is gradually shifted upward
compared to that for $g=0.0$.
Figure \ref{fig7}(b) shows a similar plot of $x(t)$ for the case of $c=10.0$, whose
period is longer than that for $c=1.0$ in Fig. \ref{fig7}(a). The time course of
$W_J(t)$ for an applied ramp force with $g=0.5$ and $\tau=100.0$
is shown by the chain (solid) curve for $c=1.0$ ($c=10.0$) in Fig. \ref{fig7}(c). 
A work induced by the applied force is $W_0=-0.123$ ($W_0=-0.101$)
for $c=1.0$ ($c=10.0$) for a given initial condition of $x_0=1.0$ and $v_0=0.0$.

Calculated WDFs for various $\tau$ are plotted in Fig. \ref{fig8}.
WDFs for $\tau=10.0$ and 100.0 have U-shaped structures, while they become the Gaussian-like 
distribution for $\tau=0.1$ and 1.0.

Figures \ref{fig9}(a), \ref{fig9}(b) and \ref{fig9}(c) show $\tau$ dependences of
$\mu$, $\sigma$ and $R$, respectively, for $c=1.0$ (dashed curves),
$c=5.0$ (dotted curves) and $c=10.0$ (solid curves).
Their $\tau$ dependences are similar to those for the van der Pol oscillator
shown in Figs. \ref{fig5}(a), \ref{fig5}(b) and \ref{fig5}(c).
We note that $\sigma$ for $c=10.0$ has a large maximum at $\tau \sim 10.0$
where $P(W)$ has the two-peak structure as shown in Fig. \ref{fig8}. 
The calculated $R$ for $c=10.0$ (solid curve)
is nearly in agreement with $\Delta F\;(=-0.125)$
for $\tau \gtrsim 20.0$ and $\tau \lesssim 0.5$,
but significantly deviates from $\Delta F$ for $0.5 \lesssim \tau \lesssim 20.0$. 
This deviation is reduced for smaller $c$ values of $c=5.0$ (dotted curve) 
and 1.0 (dashed curve), and it vanishes for $c=0.0$ 
which corresponds to a harmonic oscillator.

\section{Discussion}
\subsection{WDF of harmonic oscillators}
We have tried to elucidate the result obtained by simulations having been reported 
in the preceding section.
In a recent paper, He \cite{He03} has discussed a limit cycle of
self-excited oscillators, by using a very simple expression for
a relaxation oscillation given by
\begin{eqnarray}
x(t) &=& A \cos \omega t,
\label{eq:E1}
\end{eqnarray}
where an amplitude $A$ and frequency $\omega$
are determined as a function of a damping parameter $c$
with the use of the variational method \cite{He03}. 
It has been shown that they are given by 
$A =2.0$ and $\omega =3.8929/c + O(c^{-2})$
for the van der Pol oscillator with $c \gg 1$ \cite{He06,He03}.
Extending He's method \cite{He03}, we will study the WDF and the JE
of self-excited oscillators in the following.

Before discussing the WDF of self-excited oscillators, we briefly explain 
that of a harmonic oscillator [$c=0$ in Eq. (\ref{eq:A2})],
\begin{eqnarray}
\ddot{x}+x = f(t),
\end{eqnarray}
whose solution for the applied ramp force $f(t)$ is given by 
\begin{eqnarray}
x(t) &=& x_0 \cos t +v_0 \sin t + \int_0^t \sin(t-t') f(t')\: dt', \\
&=& x_0 \cos t +v_0 \sin t  +\frac{g(t-\sin t)}{\tau}
\hspace{1cm}\mbox{for $0 \leq t < \tau$}.
\label{eq:E15}
\end{eqnarray}
From Eq. (\ref{eq:A7}), we obtain a work for a given initial condition 
of $x_0$ and $v_0$,
\begin{eqnarray}
W_0 &=& C x_0 +D  v_0 + \phi, 
\label{eq:E16}
\end{eqnarray}
with
\begin{eqnarray}
C &=&  -\frac{g \sin \tau}{\tau}, \\
D &=& -\frac{g(1-\cos \tau)}{\tau}, \\
\phi &=& -\frac{g^2}{2}+\frac{g^2(1-\cos \tau)}{\tau^2}.
\end{eqnarray}

The WDF in Eq. (\ref{eq:A8}) is given by  
\begin{eqnarray}
P(W) &=& \frac{1}{2 \pi} \int_{-\infty}^{\infty} 
e^{i u W} \left< e^{-i u W_0} \right>_0\:du,
\label{eq:E17b}
\end{eqnarray}
with
\begin{eqnarray}
\left< e^{-i u W_0} \right>_0
&\propto &  \exp(- i u \phi) 
\int_{-\infty}^{\infty} \exp\left[ -\frac{\beta x_0^2}{2} - i u C x_0\right]\: dx_0
\int_{-\infty}^{\infty} \exp\left[ -\frac{\beta v_0^2}{2} - i u D v_0 \right]\:dv_0,
\nonumber \\
&\propto & \exp(-i u \phi) \exp\left[-\frac{(C^2+D^2)u^2}{2 \beta} \right].
\label{eq:C9}
\end{eqnarray}
A simple manipulation with Eqs. (\ref{eq:E17b}) and (\ref{eq:C9})
leads to the Gaussian WDF given by 
\begin{eqnarray}
P(W) 
&=& \frac{1}{\sqrt{2 \pi \sigma^2}}\:e^{-(W-\mu)^2/2 \sigma^2},
\label{eq:E17}
\end{eqnarray}
with
\begin{eqnarray}
\mu &=& \phi= -\frac{g^2}{2}+ \frac{g^2(1-\cos \tau)}{\tau^2},
\label{eq:E18}\\
\sigma^2 &=& \frac{2 g^2(1- \cos \tau)}{\beta \tau^2}.
\label{eq:E19}
\end{eqnarray}
The average of $e^{-\beta W}$ over $P(W)$ in Eq. (\ref{eq:A1}) is given by
\begin{eqnarray}
\left< e^{- \beta W }\right> &=& e^{-\beta(\mu-\beta \sigma^2/2)},
\end{eqnarray}
which yields 
\begin{eqnarray}
R &=& -\frac{1}{\beta} \ln \left< e^{-\beta W} \right>, 
\label{eq:E23}\\
&=& \mu-\frac{\beta \sigma^2}{2} = -\frac{g^2}{2}=\Delta F.
\label{eq:E22}
\end{eqnarray}
Equation (\ref{eq:E22}) implies that the JE holds regardless 
of a value of $\tau$ in harmonic oscillators \cite{Douarche05}-\cite{Hasegawa11b}.

\subsection{WDF of self-excited oscillators}
We now calculate the WDF of self-excited oscillators.
Taking into account Eqs. (\ref{eq:E1}) and (\ref{eq:E15}),
we have assumed that the solution of the limit cycle 
in a self-excited oscillator is given by
\begin{eqnarray}
x(t) &=& A \cos (\omega t-\theta) +\frac{g(t-\sin t)}{\tau}
\hspace{1cm}\mbox{for $0 \leq t < \tau$}.
\label{eq:E2}
\end{eqnarray}
Here $A$ and $\omega$ depend on $c$ as in Ref. \cite{He03}
and a phase $\theta$ is determined by initial conditions of $x_0$ and $v_0$, 
\begin{eqnarray}
\tan \theta = \frac{v_0}{\omega x_0},
\label{eq:E3}
\end{eqnarray}
which arises from
\begin{eqnarray}
x_0= A \cos \theta,\;\;
v_0 &=& \omega A \sin \theta.
\label{eq:E4}
\end{eqnarray}
Note that $A$ in a self-excited oscillator is assumed to
depend on $c$ but to be independent of initial condition of
$x_0$ and $v_0$, while $A$ in a harmonic oscillator
depends on them as given by $A=\sqrt{x_0^2+v_0^2}$ in Eq. (\ref{eq:E15}). 

Substituting Eq. (\ref{eq:E2}) into Eq. (\ref{eq:A7}), we obtain
a work performed by the ramp force for a given $\theta$,
\begin{eqnarray}
W_0 &=& - \left( \frac{g A}{\omega \tau} \right)
[\sin (\omega \tau-\theta) + \sin \theta]
-\frac{g^2}{2}+\frac{g^2(1-\cos \tau)}{\tau^2}.
\label{eq:E5}
\end{eqnarray}
The average of $\langle e^{-i u W_0} \rangle_0$ in Eq. (\ref{eq:E17b}) is given by
\begin{eqnarray}
\left< e^{-i u W_0} \right>_0 &\propto& 
\int_{-\infty}^{\infty} \int_{-\infty}^{\infty} 
\:e^{-\beta x_0^2/2}   \:e^{-\beta v_0^2/2} e^{- i u W_0}\: dx_0\: dv_0.
\label{eq:E7}
\end{eqnarray}
Transforming Eq. (\ref{eq:E7}) to the polar coordinate
and using Eq. (\ref{eq:E4}), we obtain
\begin{eqnarray}
\left< e^{-i u W_0} \right>_0 &\propto&
\int_0^{2 \pi} 
e^{-(\beta A^2/2)(\cos^2 \theta+ \omega^2 \sin^2 \theta)}
\:e^{i u [h(\theta)-\mu]}\: d \theta,
\label{eq:E8}
\end{eqnarray}
with
\begin{eqnarray}
h(\theta) &=& \left( \frac{g A}{\omega \tau} \right)
[\sin(\omega \tau-\theta)+\sin \theta], 
\label{eq:E9}\\
&=& 
W_d \:\cos(\theta-\delta), 
\label{eq:E10}\\
W_d &=& \sqrt{2} \:\sigma =\frac{gA \sqrt{2(1-\cos \omega \tau)}}{\omega \tau},
\label{eq:E14} \\
\tan \delta &=& \frac{(1-\cos \omega \tau)}{\sin \omega \tau },
\label{eq:E11}\\
\mu &=& -\frac{g^2}{2}+\frac{g^2(1-\cos \tau)}{\tau^2}.
\label{eq:E10b} 
\end{eqnarray}
A substitution of Eq. (\ref{eq:E8}) into Eq. (\ref{eq:E17b}) leads to 
\begin{eqnarray}
P(W) &\propto& \int_0^{2 \pi}  
e^{-(\beta A^2/2)(\cos^2 \theta+ \omega^2 \sin^2 \theta)}
\:\delta(W-\mu+h(\theta))\:d\theta, 
\end{eqnarray}
which yields the WDF given by
\begin{eqnarray}
P(W) &\simeq & \left( \frac{1}{\pi} \right)\frac{1}{\sqrt{W_d^2-(W-\mu)^2}}
\hspace{1cm}\mbox{for $-W_d+\mu < W < W_d + \mu$}.
\label{eq:E12}
\end{eqnarray}
Equation (\ref{eq:E12}) expresses the U-shaped WDF which is
divergent at two edges of $(\pm W_d+\mu)$.
Thus when $c$ is increased from zero, the WDF changes from the Gaussian
[Eq. (\ref{eq:E17})] to U-shaped non-Gaussian [Eq. (\ref{eq:E12})], 
just as shown in Fig. \ref{fig4}.

Figures \ref{fig10}(a) and \ref{fig10}(b) express $W_d$ and $\mu$, respectively,
as a function of $\tau$, which are calculated by Eqs. (\ref{eq:E14}) and (\ref{eq:E10b})
with $A=2.0$, $g=0.5$, $\omega=2 \pi/T$ and $T=19.07$ 
[Figs. \ref{fig1}(d) and \ref{fig6}(d)].
We note in Fig. \ref{fig10}(a) that the second law of thermodynamics holds because
$\mu \geq \Delta F$ ($=-0.125$).
$W_d$ has an interesting $\tau$ dependence, 
which is similar to that of $\sigma$ in harmonic oscillator 
given by Eq. (\ref{eq:E19}).
Two edges of $(\pm W_d+\mu)$ are plotted 
by solid curves in Fig. \ref{fig10}(c),
where squares (circles) express upper and lower edges of the WDF 
obtained by simulations for the van der Pol oscillator (Rayleigh oscillator).
Oscillating behaviors in upper and lower edges of the WDF obtained in simulations
are well reproduced in Fig. \ref{fig10}(c). 

By using the WDF given by Eq. (\ref{eq:E12}), 
we may evaluate $\langle e^{-\beta W}\rangle $ in Eq. (\ref{eq:A1}), 
\begin{eqnarray}
\left< e^{-\beta W} \right> &=& e^{-\beta \mu} \:I_0(\beta W_d),
\label{eq:E20}
\end{eqnarray}
where $I_n(z)$ expresses the modified Bessel function of the first kind.
From Eq. (\ref{eq:E20}), $R$ in Eq. (\ref{eq:E23}) is given by
\begin{eqnarray}
R 
&=& \mu - \frac{1}{\beta} \ln I_0(\beta W_d).
\label{eq:E21}
\end{eqnarray}
In the limit of $\tau=\infty$, we have $R=\Delta F$ because $\mu=-g^2/2=\Delta F$, 
$W_d=0.0$ and $I_0(0)=1.0$. In the opposite limit of $\tau=0.0$ where
$\mu=0.0$ and $W_d=g A$, we obtain $R=-\beta^{-1} \ln I_0(\beta g A)$
which is generally different from $\Delta F$. 
By using Eqs. (\ref{eq:E14}), (\ref{eq:E10b}) and (\ref{eq:E21})
with $A=2.0$, $g=0.5$, $\omega=2 \pi/T$ and $T=19.07$,
we have calculated $R$ which is plotted by the solid curve in Fig. \ref{fig11}.
For a comparison, we show by squares and circles, 
results of simulations for van der Pol and Rayleigh oscillators, respectively,
with $c=10.0$.
We note that the $\tau$ dependence of $R$ for $\tau \gtrsim 1.0$ obtained 
by simulations is semi-quantitatively explained by our analysis.

Our calculation, however, yields poor results for $\tau \lesssim 1.0$
where $x(t)$ 
in Eq. (\ref{eq:E2}) is not a good approximation because 
an oscillation cannot become a limit cycle for $t \sim \tau \ll T$.
Actually with decreasing $\tau$ at $\tau \lesssim 1.0$,
the WDF is changed from the U-shaped distribution
to the Gaussian distribution, as shown in Figs. \ref{fig3} and \ref{fig8}. 
Our calculation for $\tau \lesssim 1.0$ may be improved
if the WDF is phenomenologically interpolated between the Gaussian
and U-shaped distributions as given by
\begin{eqnarray}
P(W) &\propto&  \frac{p}{\sqrt{2 \pi \sigma^2}}\:e^{-(W-\mu)^2/2 \sigma^2}
+\frac{(1-p)}{\pi} \:\frac{1}{\sqrt{W_d^2-(W-\mu)^2}},
\label{eq:F1} 
\end{eqnarray}
with
\begin{eqnarray}
p &=& e^{-\tau/\tau_0},
\label{eq:F2} 
\end{eqnarray}
where $\tau_0$ denotes a parameter.
The Gaussian and U-shaped WDFs are dominant for small and large $\tau$,
respectively, and they are interpolated between small and large values of $\tau$ 
with a factor $p$.
The average of $\langle e^{-\beta W} \rangle$ is given by
\begin{eqnarray}
\left< e^{-\beta W} \right> &=& p \: e^{-\beta(\mu-\beta \sigma^2/2)}
+(1-p) \: e^{-\beta \mu} I_0(\beta W_d).
\label{eq:F3} 
\end{eqnarray}
The bold solid curve in Fig. \ref{fig11} expresses $R$ obtained 
by Eqs. (\ref{eq:E23}), (\ref{eq:F2}) and (\ref{eq:F3})
with $\tau_0=1.0$. We note that $R$ deviates from $\Delta F$
for $1.0 \lesssim \tau \lesssim 10.0$ although $R \simeq \Delta F$
for $\tau \ll 1.0$ and $\tau \gg 10$, as shown by simulations.

\section{Conclusion}
Studying the JE in van der Pol and Rayleigh oscillators to which
a ramp force with a duration $\tau$ is applied, 
we have obtained the following results:

\noindent
(i) The JE nearly holds in a fairly wide range of $\tau$ including
transient ($\tau \rightarrow 0$) and quasi-stationary forces ($\tau \rightarrow \infty$),
although the JE is not rigorously satisfied  \cite{Cuendet06,Cuendet06b,Paschinger06},

\noindent
(ii) The WDF has the U-shaped structure for a large damping parameter, and

\noindent
(iii) The $\tau$ dependence of 
$R$ $\left(=- k_B T \ln \left< e^{-\beta W} \right> \right)$
[Eq. (\ref{eq:A10})] obtained 
by our simulations may be semi-quantitatively accounted for 
by our analysis with a simple expression of $x(t)$ for a limit cycle whose amplitude
is assumed to be determined by a damping parameter but not sensitive to initial conditions.

\noindent 
The item (i) is in contrast with results of NH oscillators 
where JE holds \cite{Cuendet06,Cuendet06b,Paschinger06}.
Derivations of the JE require that
the equilibrium canonical distribution in non-Hamiltonian systems
should satisfy the condition given 
by Eqs. (\ref{eq:G6}) and (\ref{eq:G6b}) \cite{Paschinger06}.
Van der Pol and Rayleigh oscillators do not meet the condition,
while it is held in the NH oscillator \cite{Cuendet06,Cuendet06b,Paschinger06}.
Although our simple analysis in the item (iii) may 
explain essential features of van der Pol and Rayleigh models,
a development of more advanced theory is desirable
for a better understanding of their properties. 
It would be interesting to examine our result by experiments, for example,
by electrical circuits consisting of nonlinear elements.

\begin{acknowledgments}
This work is partly supported by a Grant-in-Aid for Scientific Research from 
Ministry of Education, Culture, Sports, Science and Technology of Japan.  
\end{acknowledgments}

\appendix*
\section{A. Energy, heat and work of the van der Pol oscillator}
\renewcommand{\theequation}{A\arabic{equation}}
\setcounter{equation}{0}

In this Appendix we will present some model calculations of thermodynamical quantities 
such as the energy and heat in the van der Pol oscillator, 
which are evaluated both by single and multiple runs of simulations.
Figures \ref{fig12}(a), \ref{fig12}(b), \ref{fig12}(c) and \ref{fig12}(d) show 
$x(t)$, $U(t)$, $dQ(t)/dt$ and $W_J(t)$, respectively, of the van der Pol oscillator with $c=1.0$
for applied ramp forces with $g=0.0$ (dashed curve) and $g=0.5$ (solid curve) for $\tau=10.0$
calculated by single runs with initial conditions of $x_0=1.0$ and $v_0=1.0$ $[U(0)=1.0]$.
The period of the oscillation with the applied ramp force 
with $g=0.5$ is gradually increased compared to that with $g=0.0$. 
We obtain $U(t)-U(0)=Q(t)$ for $g=0.0$ where $W_J(0)=W_c(t)=0.0$ in Eq. (\ref{eq:D1}). 
The heat (energy) flows from an environment to the oscillator for $dQ(t)/dt > 0$,
and for $dQ(t)/dt < 0$ the heat (energy) flow is reversed.
Periodic energy exchanges are realized between the oscillator and environment
in the limit cycle.
For an applied ramp force with $g=0.5$, $W_J(t)$ is time dependent at $0 \leq t < 10.0$ 
and it becomes constant ($=-0.191$) at $t \geq 10.0$ where $\dot{f}(t)=0$, 
as shown in Fig. \ref{fig12}(d).

Figures \ref{fig12}(e)- \ref{fig12}(h) show similar plots of
relevant thermodynamical quantities in the van der Pol oscillator with 
a larger damping constant of $c=10.0$ which are calculated also by single runs with the same
initial condition of $x_0=1.0$ and $v_0=1.0$.
The period of relaxation oscillation of $x(t)$ for $c=10.0$ 
is larger than that for $c=1.0$. 
Time dependences of $U(t)$ and $dQ(t)/dt$ for $c=10.0$ become 
much significant than those for $c=1.0$: note that scales of ordinates 
in Figs. \ref{fig12}(f) and \ref{fig12}(g) are much larger than those 
in Figs. \ref{fig12}(b) and \ref{fig12}(c).
We obtain a positive $W_J(t)$ $(=0.551)$ at $t \geq 10.0$ in Fig. \ref{fig12}(h). 

Related thermodynamical quantities averaged over 100,000 runs 
with canonically distributed initial states of $\{ x_0 \}$ and $\{ v_0\}$ with $k_B T=1.0$ 
[Eqs. (\ref{eq:A8b}) and (\ref{eq:A9})] are plotted in Fig. \ref{fig13}.
Figures \ref{fig13}(a), \ref{fig13}(b), \ref{fig13}(c) and \ref{fig13}(d) show 
$\langle x(t) \rangle_0$, $\langle U(t) \rangle_0$, $\langle dQ(t)/dt \rangle_0$  
and $\langle W_J(t) \rangle_0$, respectively, of the van der Pol oscillator with $c=1.0$ 
for applied forces of $g=0.0$ (dashed curves) and $g=0.5$ (solid curves) with $\tau=10.0$.
We note in Fig. \ref{fig13}(a) that although $\langle x \rangle_0=0$ for $g=0.0$,
$\langle x \rangle_0$ for $g=0.5$ at $t \geq 10.0$ 
expresses a small limit-cycle oscillation superposed on a constant of 0.5.
This is because random initial states are effectively biased by an applied force.
For $g=0.0$, $\langle U(t) \rangle_0 =1.0$ at $t=0.0$
and it becomes about 2.0 at $t \gtrsim 5.0$ which is determined by amplitudes of 
$x(t)$ and $v(t)$ in the limit cycle, as shown in Fig. \ref{fig13}(b).
It is noted that even for $g=0.0$, we obtain 
$\langle U(t) \rangle_0- \langle U(0) \rangle_0 \neq 0.0$ 
which is due to finite dissipative contributions of $\langle dQ/dt \rangle_0$ $(\neq 0.0)$
between an oscillator and environment.
Comparing Fig. \ref{fig13}(b) with Fig. \ref{fig12}(b), 
we note that magnitudes of $\langle U(t) \rangle_0$ become much smaller than those of $U(t)$
for a single run.
Owing to an applied ramp force, $\langle U(t) \rangle_0$ for $g=0.5$
is lower than that for $g=0.0$. We obtain $\langle W_J \rangle_0= -0.110$ at $t \geq 10.0$
as shown in Fig. \ref{fig13}(d).

Similar plots of relevant thermodynamical quantities for the van der Pol oscillator 
with a larger $c=10.0$ are presented in Figs. \ref{fig13}(e)-\ref{fig13}(h). The initial averaged energy 
of $\langle U(t) \rangle_0=1.0$ at $t=0.0$ is increased to about $2.4 \sim 3.6$ at $t \gtrsim 5.0$
for $g=0.0$ in Fig. \ref{fig13}(f). This increase in $\langle U(t) \rangle_0$ is due to
energy supplies from environment to the oscillator 
which are rapidly accomplished at $0 \leq t \lesssim 1.0$ both for $g=0.0$ and 0.5
as shown in  Figs. \ref{fig13}(f) and \ref{fig13}(g).
Figure \ref{fig13}(h) shows $\langle W_J \rangle_0= -0.092$ at $t \geq 10.0$,
which is in contrast to a positive $W_J=0.551$ for a single run shown in Fig. \ref{fig12}(h).


\newpage

\begin{figure}
\begin{center}
\end{center}
\caption{
(Color online)
(a) $x(t)$, (b) $v(t)$ and (c) a parametric plot of 
$x(t)$ vs. $v(t)$ in the van der Pol oscillator with $c=10.0$
for $f(t)=0.0$ with an initial condition of $x_0=1.0$ and $v_0=0.0$. 
(d) The $c$ dependence of a period $T$ with constant forces 
of $f(t)=0.0$ (dashed curve) and $f(t)=0.5$ (solid curve).
}
\label{fig1}
\end{figure}

\begin{figure}
\begin{center}
\end{center}
\caption{
(Color online) 
Time courses of $x(t)$ of the van der Pol oscillator with (a) $c=1.0$ and  
(b) $c=10.0$ for ramp forces with $g=0.0$ (dashed curve) and $g=0.5$ (solid curve) for $\tau=100.0$.
(c) $W_J(t)$ with $c=1.0$ (chain curve) and $c=10.0$ (solid curve) for $g=0.5$ and $\tau=100.0$.
An applied ramp force $f(t)$ is plotted by dotted curves in (a) and (b).
Simulations are performed by single runs with initial conditions of $x_0=1.0$ and $v_0=0.0$.
}
\label{fig2}
\end{figure}

\begin{figure}
\begin{center}
\end{center}
\caption{
(Color online) 
$P(W)$ for $\tau=0.1$, 1.0, 10.0 and 100.0 with $c=10.0$ and $g=0.5$ 
in the van der Pol oscillator, $P(W)$ for $\tau=10.0$ and 100.0 being multiplied 
by factor of 1/2 and 1/20, respectively.
}
\label{fig3}
\end{figure}

\begin{figure}
\begin{center}
\end{center}
\caption{
(Color online) 
$P(W)$ for $c=1.0$, 2.0, 5.0 and 10.0 with 
$\tau=1.0$ and $g=0.5$ in the van der Pol oscillator.
}
\label{fig4}
\end{figure}

\begin{figure}
\begin{center}
\end{center}
\caption{
(Color online) 
The $\tau$ dependence of (a) $\mu$, (b) $\sigma$ and (c) $R$ 
in the van der Pol oscillator with $c=1.0$ (dashed curve), 5.0 (dotted curve) 
and 10.0 (solid curve) for ramp forces with $g=0.5$, 
the arrow along the right ordinate in (c) expressing $\Delta F$.
The JE is expressed by $R=\Delta F$ $(=-0.125)$.
}
\label{fig5}
\end{figure}

\begin{figure}
\begin{center}
\end{center}
\caption{
(Color online) 
(a) $x(t)$, (b) $v(t)$ and (c) a parametric plot of $x(t)$ vs. $v(t)$ 
in the Rayleigh oscillator with $c=10.0$ for $f(t)=0.0$
with the initial condition of $x_0=1.0$ and $v_0=0.0$. 
(d) The $c$ dependence of a period $T$ 
with constant forces of $f(t)=0.0$ (dashed curve) and $f(t)=0.5$ (solid curve).
}
\label{fig6}
\end{figure}

\begin{figure}
\begin{center}
\end{center}
\caption{
(Color online) 
Time courses of $x(t)$ of the Rayleigh oscillator with (a) $c=1.0$ and 
(b) $c=10.0$ for ramp forces with $g=0.0$ (dashed curve) and $g=0.5$ (solid curve) for $\tau=100.0$. 
(c) $W_J(t)$ with $c=1.0$ (chain curve) and $c=10.0$ (solid curve) for $g=0.5$ with $\tau=100.0$.
An applied ramp force $f(t)$ is plotted by dotted curves in (a) and (b).
Simulations are performed by single runs with initial conditions of $x_0=1.0$ and $v_0=0.0$.
}
\label{fig7}
\end{figure}

\begin{figure}
\begin{center}
\end{center}
\caption{
(Color online) 
$P(W)$ for $\tau=0.1$, 1.0, 10.0 and 100.0 with $c=10.0$ and $g=0.5$ 
in the Rayleigh oscillator, $P(W)$ for $\tau=10.0$ and 100.0 being multiplied 
by factors of 1/10 and 1/5, respectively, and that for $\tau=1.0$ being shifted
upward by 0.2.
}
\label{fig8}
\end{figure}

\begin{figure}
\begin{center}
\end{center}
\caption{
(Color online) 
The $\tau$ dependence of (a) $\mu$, (b) $\sigma$ and (c) $R$
in the Rayleigh  oscillator with $c=1.0$ (dashed curve), 5.0 (dotted curve) 
and 10.0 (solid curve) for ramp forces with $g=0.5$, 
the arrow along the right ordinate in (c) expressing $\Delta F$ ($=-0.125$).
}
\label{fig9}
\end{figure}

\begin{figure}
\begin{center}
\end{center}
\caption{
(Color online) 
(a) $W_d$ and (b) $\mu$ as a function of $\tau$
calculated by Eqs. (\ref{eq:E14}) and (\ref{eq:E10b}) with
$A=2.0$, $g=0.5$, $\omega=2 \pi/T$ and $T=19.07$.
(c) The $\tau$ dependence of $(\pm W_d+\mu)$ (solid curves) and 
that of upper and lower edges of the WDF obtained by simulations
for van der Pol (squares) and Rayleigh oscillators (circles) with $c=10.0$,
dashed and chain curves being plotted only for a guide of eye (see text).
}
\label{fig10}
\end{figure}

\begin{figure}
\begin{center}
\end{center}
\caption{
(Color online) 
The $\tau$ dependence of $R$ calculated 
by Eq. (\ref{eq:E21}) (solid curves) and 
Eq. (\ref{eq:F3}) (interpolation: bold solid curve) with
$A=2.0$, $g=0.5$, $\omega=2 \pi/T$ and $T=19.07$, and those
for van der Pol (squares) and Rayleigh oscillators (circles) with $c=10.0$
obtained by simulations,
dashed and chain curves being plotted only for a guide of eye.
The arrow along the right ordinate expresses $\Delta F$ ($=-0.125$).
}
\label{fig11}
\end{figure}

\begin{figure}
\begin{center}
\end{center}
\caption{
(Color online) 
(a) $x(t)$, (b) $U(t)$, (c) $dQ(t)/dt$ and (d) $W_J(t)$ in the van der Pol oscillator with $c=1.0$,
and (e) $x(t)$, (f) $U(t)$, (g) $dQ(t)/dt$ and (h) $W_J(t)$ with $c=10.0$
for applied ramp forces with $g=0.0$ (dashed curves) and $g=0.5$ (solid curves)
($\tau=10.0$) evaluated by single runs with the initial condition of $x_0=1.0$ and $v_0=1.0$
yielding $U(0)=1.0$.
}
\label{fig12}
\end{figure}

\begin{figure}
\begin{center}
\end{center}
\caption{
(Color online) 
(a) $\langle x(t) \rangle_0$, (b) $\langle U(t) \rangle_0$, (c) $\langle dQ(t)/dt \rangle_0$ 
and (d) $\langle W_J(t) \rangle_0$ in the van der Pol oscillator with $c=1.0$,
and (e) $\langle U(t) \rangle_0$, (f) $\langle Q(t) \rangle_0$, (g) $\langle dQ(t)/dt \rangle_0$
and (h) $\langle W_J(t) \rangle_0$ with $c=10.0$
for applied ramp forces with $g=0.0$ (dashed curves) and $g=0.5$ (solid curves)
($\tau=10.0$) averaged over $100,000$ runs with $k_B T=1.0$ $(=\langle U(0) \rangle_0)$.
Results for $g=0.5$ in (f) and (g) are shifted upward by five and ten, respectively, 
for a clarity of figures.
}
\label{fig13}
\end{figure}

\end{document}